\documentclass[a4paper,aps,prd,10pt,preprintnumbers,showpacs,twocolumn,superscriptaddress,nofootinbib,amsmath,amssymb]{revtex4-1}
\usepackage[utf8]{inputenc}
\usepackage[T1]{fontenc}
\usepackage{cmap}
\usepackage{graphicx}

\begin{document}
\title{Emergence of Negative Mass in General Relativity}

\author{Chen-Hao Hao}
\author{Long-Xing Huang}
\author{Xin Su}
\author{Yong-Qiang Wang}\email{yqwang@lzu.edu.cn, corresponding author}
\affiliation{ $^{}$Lanzhou Center for Theoretical Physics, Key Laboratory of Theoretical Physics of Gansu Province,
	School of Physical Science and Technology, Lanzhou University, Lanzhou 730000, China\\
	$^{2}$Institute of Theoretical Physics $\&$ Research Center of Gravitation, Lanzhou University, Lanzhou 730000, China}

\begin{abstract}
We develop a symmetric traversable wormhole model, integrating Einstein's gravitational coupling phantom field and a nonlinear electromagnetic field. This work indicates the emergence of negative ADM mass within a specific parameter range, coinciding with distinct alterations in the wormhole's spacetime properties. Despite violating the Null Energy Condition (NEC) and other energy conditions, the solution exhibits unique characteristics in certain energy-momentum tensor components, potentially accounting for the manifestation of negative mass.
\end{abstract}
\maketitle

\section{Introduction}
In the early 1980s, the Positive Mass Theorem was established following the validation of the positive mass conjecture by R. Schoen, S. T. Yau, and E. Witten \cite{Schon:1979rg, Witten:1981mf}. This theorem asserts that for an isolated system if the material distribution complies with the Dominant Energy Condition (DEC), the ADM mass is ensured to be non-negative. The theorem's proof is mathematically complex and relies heavily on the DEC, denoted as  $\rho \geq\left|p_{i}\right|(i=1,2,3)$. Other energy conditions posit that various linear combinations of the stress-energy tensor components should be positive, aligning with intuitive understanding. Moreover, by assuming various positive energy-momentum tensor components, it is possible to prove the singularity theorem (dependent on SEC), and the topological censorship theorem (dependent on NEC), among others \cite{Hawking:1970zqf, Hawking:1991nk, Friedman:2006tea}. Over the years, perspectives on energy conditions have evolved. Through the construction of various models that violate energy conditions, researchers have uncovered a wealth of novel physical phenomena.

Among the myriad of exotic physical phenomena resulting from energy condition violations, the concept of traversable wormholes emerges as particularly intriguing. The wormhole theory, initially proposed by A. Einstein and N. Rosen in 1935 and referred to as the ``Einstein-Rosen Bridge'' \cite{Einstein:1935tc}, is non-traversable. It wasn't until 1973 that H. Ellis and K. Bronnikov independently introduced a model for traversable wormholes \cite{Ellis:1973yv, Bronnikov:1973fh}. Subsequently, in 1988, M. Morris and K. Thorne conducted significant research on this type of traversable wormhole, contributing to its explicit study \cite{Morris:1988cz}. The construction of such a traversable wormhole requires exotic matter with a negative energy density, primarily violating the Null Energy Condition (NEC), and generally all other energy conditions \cite{Visser:1989kh}. This type of traversable wormhole, supported by exotic matter, signifies not only the potential for large-scale spatial travel but also the prospect of time travel \cite{Morris:1988tu}. In particular, it has been proposed that the universes on both sides connected by traversable wormholes can be enriched in positive and negative masses respectively \cite{Visser:1995cc}. Indeed, within the framework of general relativity, the emergence of negative mass is not precluded if energy conditions are not imposed a priori \cite{Bondi:1957zz}. Some propose that the manifestation of negative mass matter is inevitable due to symmetry requirements and could serve as a potential candidate for dark energy, contributing to the accelerated expansion of the universe \cite{Farnes:2017gbf}.

Current research about negative mass primarily involves the direct introduction of negative mass substances to investigate their properties and associated effects. However, no existing model can directly and specifically yield a negative ADM mass. In this study, we present a symmetric traversable wormhole model that integrates Einstein's gravitational coupling phantom field and a nonlinear electromagnetic field, leading to the emergence of the negative ADM mass within a specific parameter range. The magnetic monopoles that we introduced here originated from the Bardeen black hole and were proposed as a means to eliminate the singularity within the Schwarzschild black hole \cite{Bardeen1}.

\section{Basic equations}
We consider the Einstein-Hilbert action including the Lagrangian for a nonlinear electromagnetic field and the phantom scalar field, the action is given by

\begin{equation}\label{action}
 S=\int\sqrt{-g}d^4x\left(\frac{R}{2\kappa}+\mathcal{L}_{1}+\mathcal{L}_{2}\right),
\end{equation}
the term
$\mathcal{L}_{1}$ and $\mathcal{L}_{2}$ are  the Lagrangians defined by with

\begin{eqnarray}
\mathcal{L}_{1} &= &- \frac{ 3}{ 2 s } \left( \frac{ \sqrt{2 q^2 {\cal F}}}{ 1 + \sqrt{ 2 q^2 {\cal F}}} \right)^{\frac{5}{2}}, \ \mathcal{L}_{2}=\nabla_a\Phi\nabla^a\Phi \ .
\end{eqnarray}
Where $R$ is the scalar curvature, the $\kappa = 2$, and  $\mathcal{L}_{1}$ is a function of ${\cal F} = \frac{1}{4}F_{ab} F^{ab}$ with the electromagnetic field strength $ F_{ab} =  \partial_{a} A_{ b} - \partial_{b} A_{ a}$, in which $A$ is  the electromagnetic field, and the $\Phi$ represent the phantom field. The constants $q$ and $s$ are two independent parameters, where $q$ represents the magnetic charge. By varying the action (1) with
respect to the metric, electromagnetic field, and phantom field respectively,  we can obtain the following equations of motion:
\begin{equation}
  \label{eq:EKG1}
R_{\mu\nu}-\frac{1}{2}g_{\mu\nu}R=\kappa T_{\mu\nu}, \
\bigtriangledown_{a} \left(\frac{ \partial {\cal L}^{(1)}}{ \partial {\cal F}}  F^{a b}\right)=0, \
\Box\Phi=0,
\end{equation}
with stress-energy tensor:
\begin{equation}
T_{\mu\nu} = g_{\mu\nu}({{\cal L}}_1+{{\cal L}}_2)
-2 \frac{\partial ({{\cal L}}_1+{{\cal L}}_2)}{\partial g^{\mu\nu}} \ .
\end{equation}

The general static spherically symmetric solution with a wormhole is given by the
following line element \cite{Dzhunushaliev:2014bya}:
\begin{equation}  \label{line_element1}
 ds^2 = -e^{B}  dt^2 +C e^{-B}   \left[ d r^2 + h (d \theta^2+\sin^2 \theta d\varphi^2)   \right]\,,
\end{equation}
here $B$ and $C$ are functions of  radial coordinate $r$,  $h=r^2+r_0^2$ with  the throat parameter  $r_0$,
 and $r$  ranges from positive infinity to negative infinity. In addition, we use the
following ansatzes of the electromagnetic field and the phantom field:
\begin{equation}\label{equ11}
   A= q \cos(\theta)d\varphi,\;\;\; \Phi=\phi(r),
\end{equation}

Substituting the above ansatzes (5) and (6) into the field equations (3), we can get the following equations:
\begin{equation}
\begin{split}
&\frac{C}{2}(-\frac{6 \kappa e^{\frac{3}{2} B} q^{5} C^{2}}{S(e^{B} q^{2}+h C)^{\frac{5}{2}}}+(-\frac{6 r}{h}+B^{\prime}) C^{\prime} \\&  + C'^{2}
+2 C(\frac{2 r B^{\prime}}{h}+B^{\prime \prime})-2 C^{\prime \prime}) = 0,
\end{split}
\end{equation}
\begin{equation}
\begin{split}
\kappa \frac{2 C^{3} h e^{\frac{3 B}{2}} q^{5}\left(6 e^{B} q^{2}-\frac{3}{2} h C\right)}{s\left(e^{B} q^{2}+h C\right)^{\frac{7}{2}}}+3 r C^{\prime}-C^{\prime 2}+2 C C^{\prime \prime} h = 0,
\end{split}
\end{equation}
\begin{equation}
\phi' = \frac{\sqrt{\cal D}}{h \sqrt{C}}\ ,
\end{equation}
the $\cal D$ is a constant that represents the scalar charge of the phantom field and can be used to check the accuracy of numerical calculations. Its value as a function of parameter $q$ should be the same at different locations while fixing throat size $r_0$. We give the expression of scalar charge $\cal D$ by taking the above Eq. (9) into the Einstein's field equation components:
\begin{equation}
{\cal D}=
\frac{2  h^{2} C}{\kappa} ((\frac{r_{0}^{2}}{h^{2}}-\frac{\frac{3}{2} e^{\frac{3 B}{2}} \kappa q^{5} C}{s\left(e^{B} q^{2}+h C\right)^{5 / 2}}+\frac{1}{4} B^{\prime 2}-\frac{r C^{\prime}}{h C}-\frac{C^{\prime 2}}{4 C^{2}})).
\label{eqD2}
\end{equation}

Solving these OED equations numerically, we can get all information about metric functions $B$, $C$, and the scalar charge $\cal D$ of the phantom field. We give the boundary conditions of the two functions B and C at infinity ($r \rightarrow \infty$):
\begin{equation}
B = 0, \ C = 1, \ \partial_r C= 0.
\end{equation}

\section{Numerical results}
In this article, we fix the parameter $s$ to 1 without loss of generality. For other $s$ values, there are consistent conclusions. To facilitate numerical calculations, we can compactify   the radial coordinate $r$ using the change of variables
$x= \frac{2}{\pi}\arctan(r)$.

Two independent asymptotically flat universes are symmetric about the wormhole throat and the ADM mass can be obtained from the Komar expression
\begin{equation}
M= \int_{\Sigma} R_{\mu \nu} n^{\mu} \xi^{\nu} d V .
\end{equation}

\begin{figure}[htbp]
\centering
\includegraphics[width=0.49\linewidth]{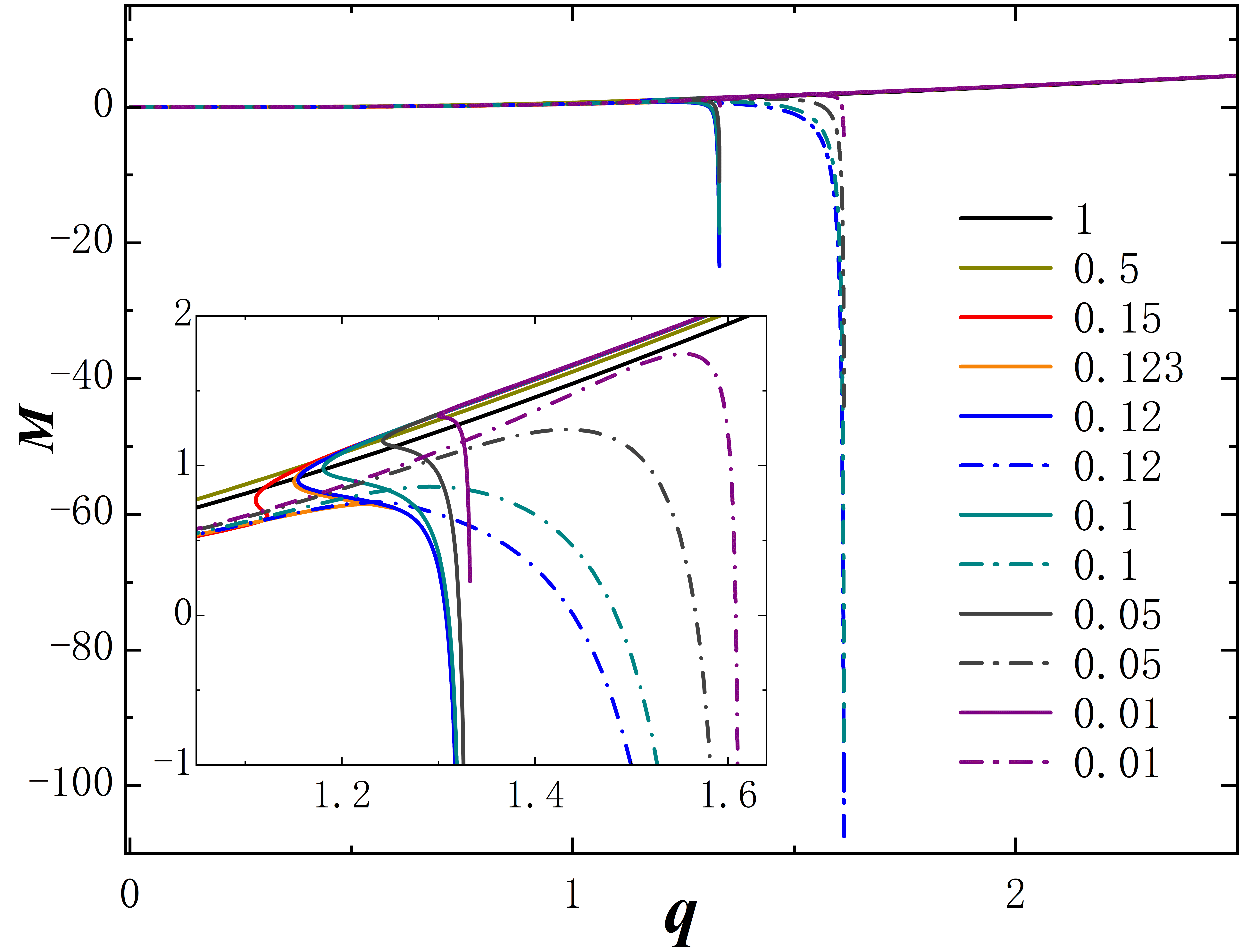}
\includegraphics[width=0.49\linewidth]{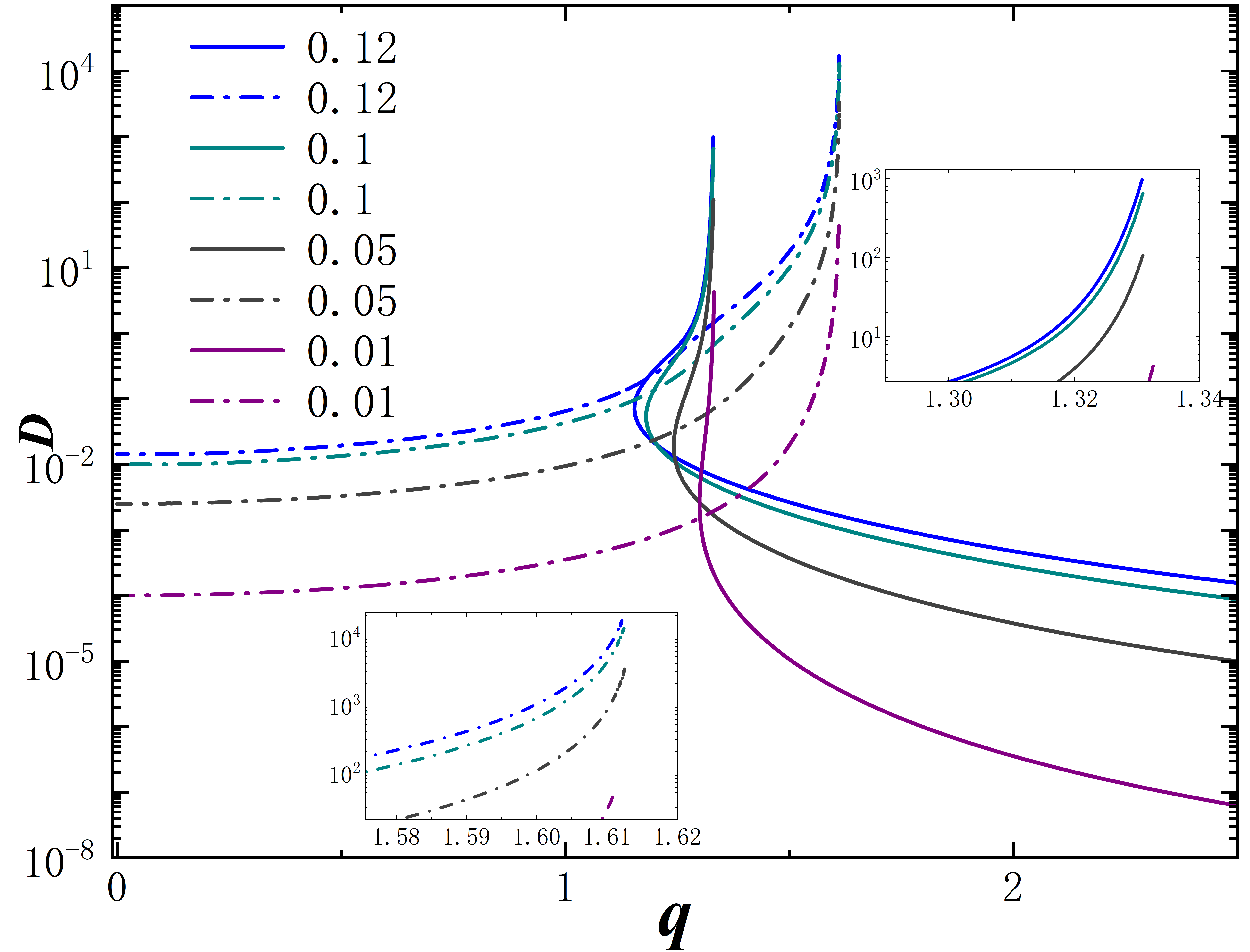}
\caption{The left-hand image depicts the functional relationship of $M$ concerning $q$, with the parameter $s$ fixed at 1 and multiple distinct $r_0$ values considered. The subfigure provides a detailed view of the mass curve within a specific $q$ range. The figure on the right shows the function graph of $\cal D$ concerning $q$ in the two solutions with different $r_0$. In both figures, the solid line represents the first solution, and the dotted represents the second.}
\label{phase1}
\end{figure}

By varying the throat size $r_0$, we derived a series of functional relationships between mass $M$ and the parameter $q$ as depicted in Fig. \ref{phase1}. We only show images within a certain range of $q$ values and because the wormhole is symmetric, only the ADM mass of one side of the universe needs to be studied. For large values of $r_0$, $M$ exhibits a monotonic increase with $q$, suggesting that a larger $q$ value corresponds to a greater contribution from the magnetic monopole, thereby increasing the mass. When $r_0 < 0.2$, the monotonic increase generally persists, but within a certain small range of $q$, the $M$ curve exhibits a ``fluctuation'', the smaller $r_0$ is, the ``fluctuation'' appears where $q$ is larger. This is attributed to the emergence of solution branches, and within this small branch range, $M$ increases as $q$ decreases. A more significant feature emerges when $r_0$ falls below the critical value of 0.123, at which point two distinct solutions appear. One solution indicates that $M$ initially decreases as $q$ reduces, then takes a sharp turn, and within a very short range, it drops to a negative value as $q$ increases. The other solution demonstrates that $M$ initially increases as $q$ rises from 0. However, after reaching a certain value, $M$ sharply declines with increasing $q$, eventually dropping to a significantly negative value.  Meanwhile, we note that as $r_0$ decreases further, the absolute value of the negative mass values attainable by both solutions also decreases. Notably, when $r_0$ is 0.01, one solution fails to yield a negative mass, and if $r_0$ is smaller, neither solution can give the negative ADM mass.

The scalar charge $\cal D$ serves as a direct measure of the phantom field quantity, and we show the graph of $\cal D$ as a function of $q$ within different values of $r_0$ in Fig. \ref{phase1}. Within the $q$ values range where the negative mass solution is observed, $\cal D$ also escalates to very large, indicating a high level of phantom field. The Komar integration of highly concentrated negative energy density may result in negative ADM mass.

\begin{figure}[htbp]
\centering
\includegraphics[width=0.49\linewidth]{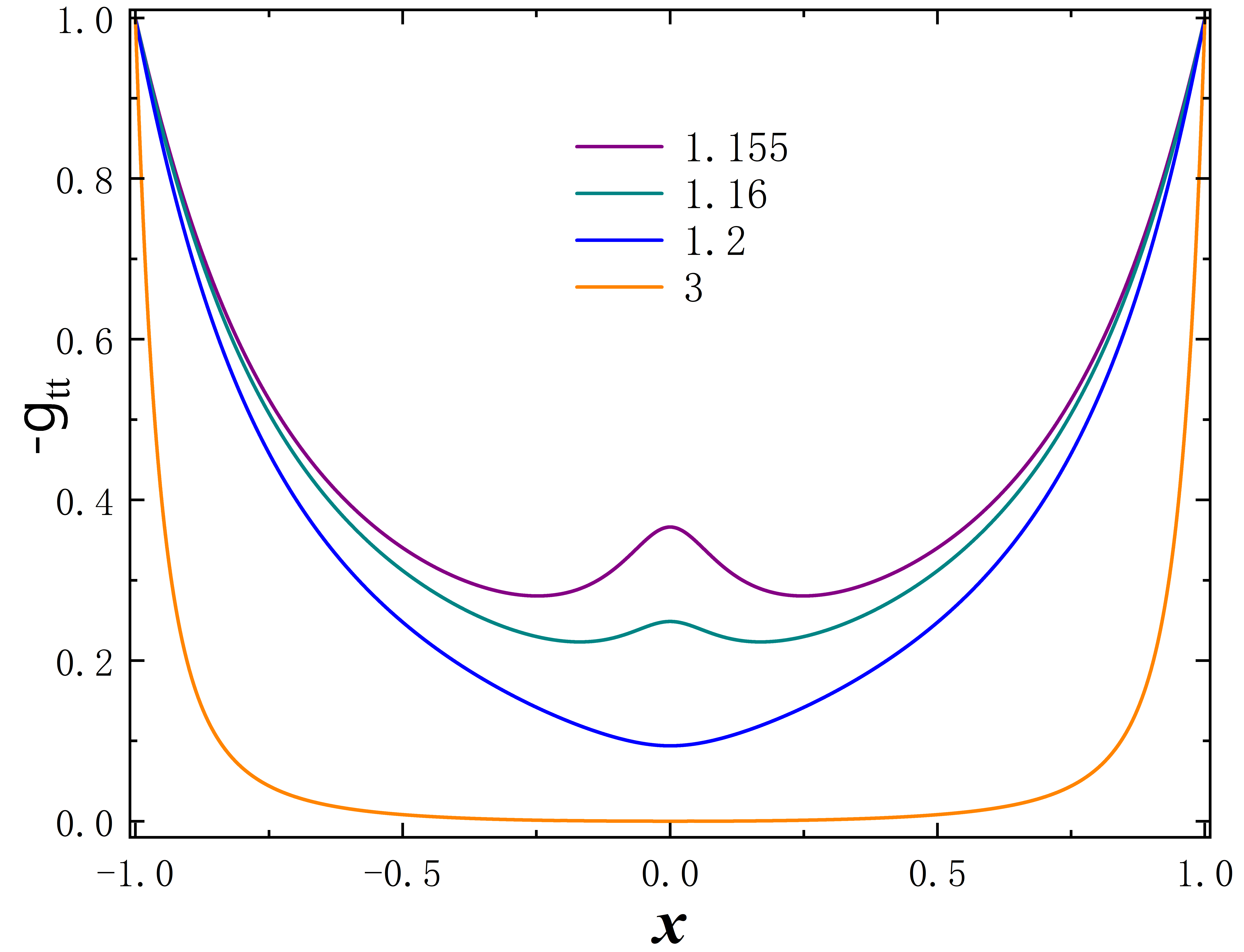}
\includegraphics[width=0.49\linewidth]{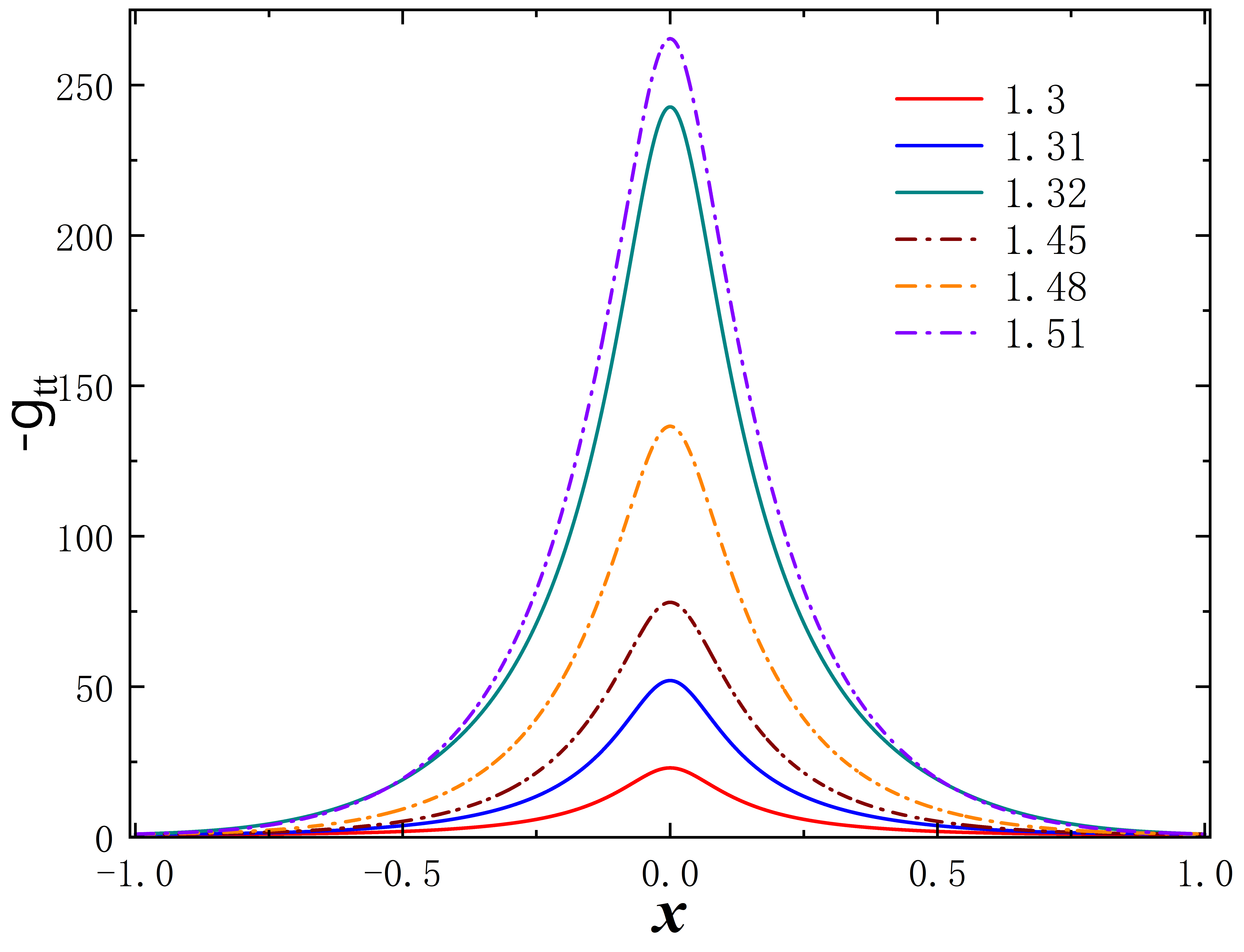}
\caption{The functional relationship of $-g_{tt}$ concerning $x$, with the parameter $s$ fixed at 1, $r_0$ fixed at 0.12 and $q$ with some different values. The left image corresponds to the scenario of positive mass, while the right image pertains to negative mass. In both figures, the solid line represents the first solution, and the dotted lines represent the second.}
\label{phase2}
\end{figure}

The metric exhibits unusual characteristics that emerge in association with the negative ADM mass. For illustrative purposes, we choose $r_0$ as 0.12 without loss of generality. In the scenario of positive mass at the first solution, the minimum value of the metric $-g_{tt}$ (at $x = 0$) asymptotically approaches zero without attaining it as $q$ increases, signifying the formation of an extreme quasi-horizon. As $q$ approaches the negative mass region, the value of $-g_{tt}$ generally increases, and its minimum value is no longer situated at $x = 0$. When $q$ is incorporated into the negative mass solution, the value of $-g_{tt}$ further increases as the magnitude of the negative mass intensifies. This is particularly evident at $x = 0$, the throat of the wormhole, where all $-g_{tt}$ values exceed 1. If adventurers were to traverse this wormhole, they would directly observe a dramatic increase in the blue shift effect as $q$ escalates. Good luck to them. In another solution where negative mass occurs, the situation is similar. As $q$ extends to the endpoint of the negative mass solution, which corresponds to the maximum negative mass, the $-g_{tt}$ value near the throat of the wormhole escalates to a significantly high order of magnitude. At the same time, the Kretschmann scalar also surges, making further numerical computation unfeasible. At this time or under conditions where $-g_{tt}$ is exceedingly close to 0 with a quasi-horizon emerges, the wormhole transitions into untraversable. Fig. \ref{phase2} illustrates the variation of $-g_{tt}$ for $x$, here $r_0$ is held constant at 0.12, with different $q$. The solid line denotes the first solution, while the dotted line signifies the second solution.

\begin{figure}[htbp]
\centering
\includegraphics[width=0.49\linewidth]{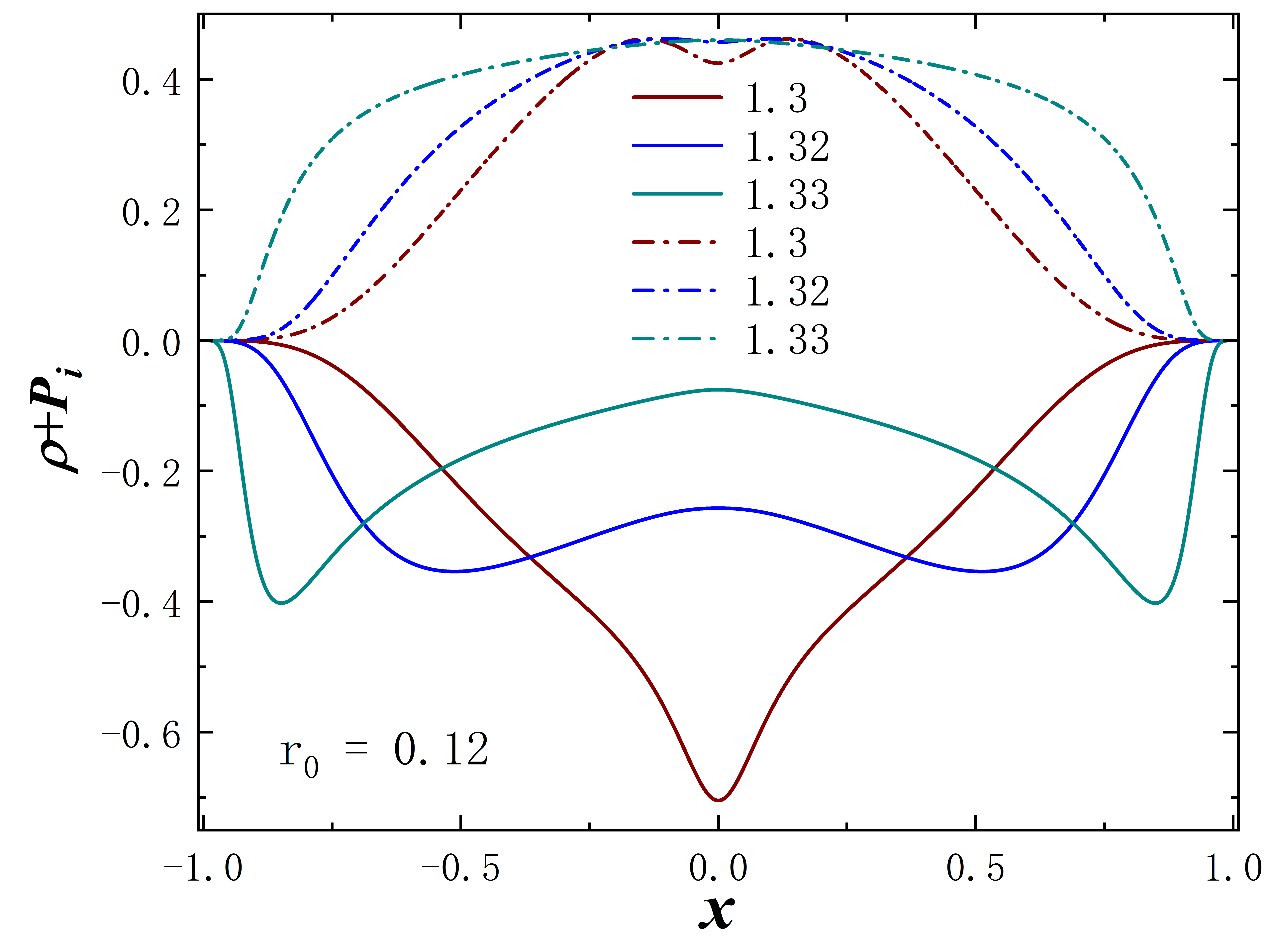}
\includegraphics[width=0.49\linewidth]{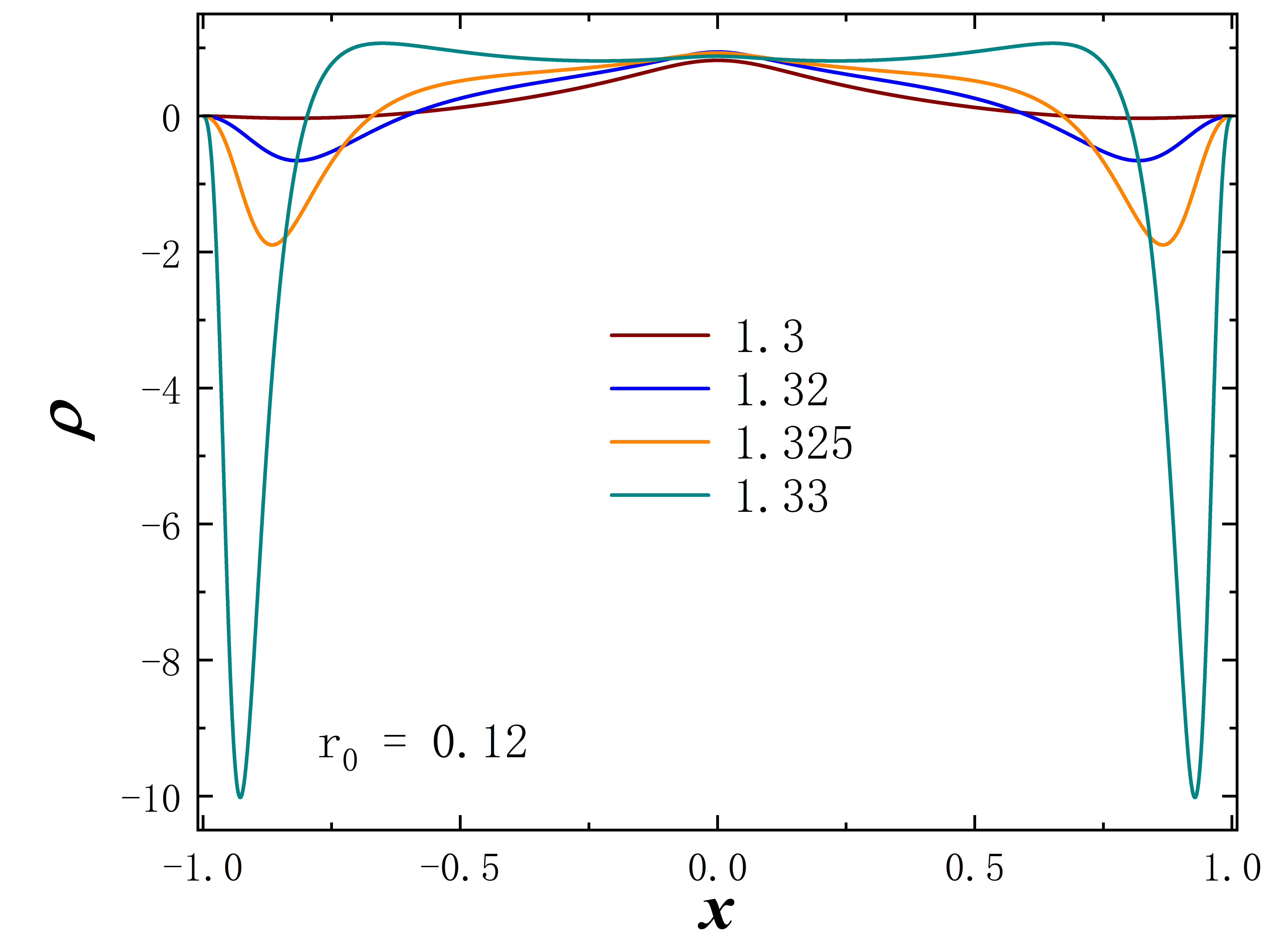}
\includegraphics[width=0.49\linewidth]{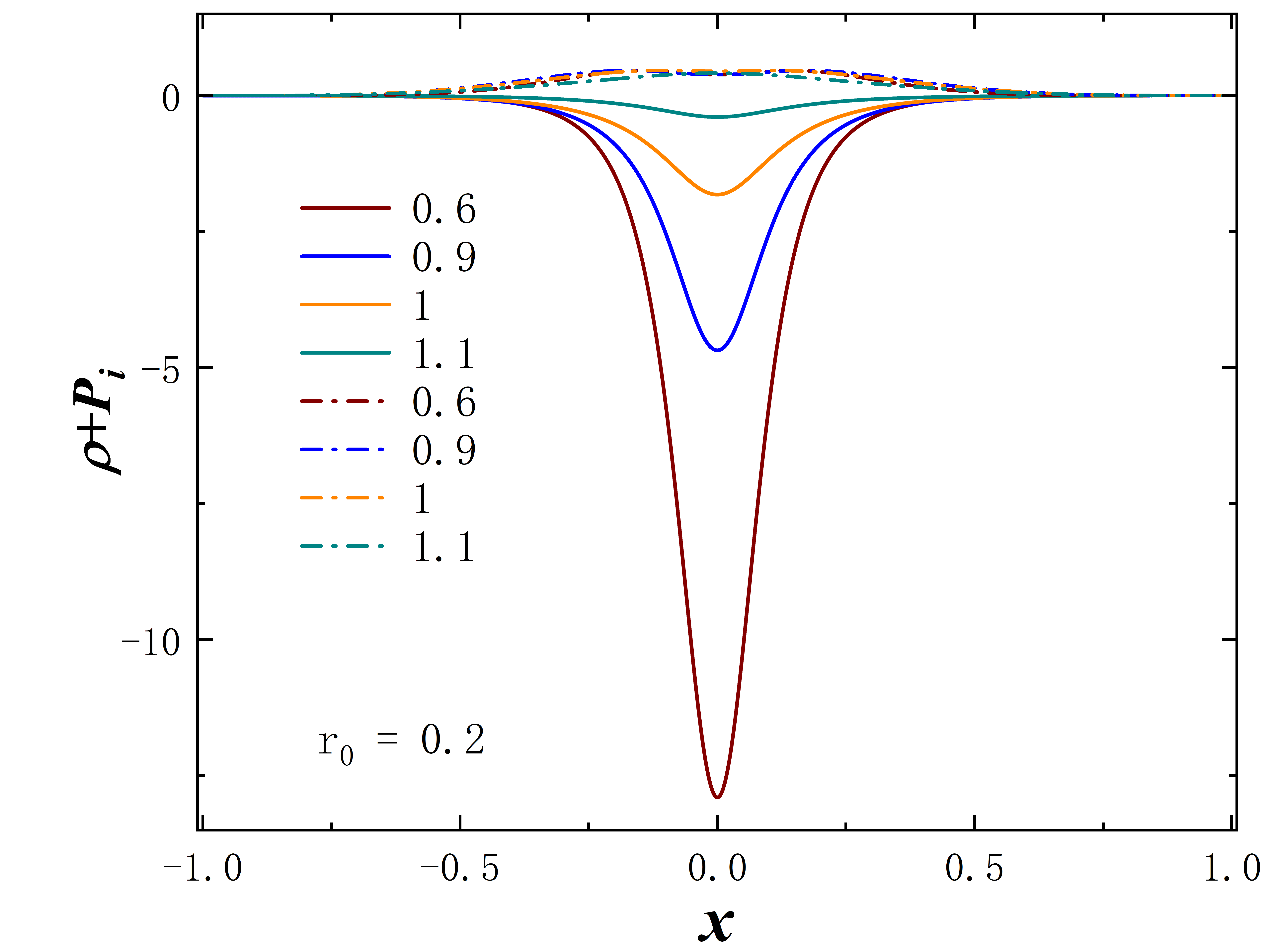}
\includegraphics[width=0.49\linewidth]{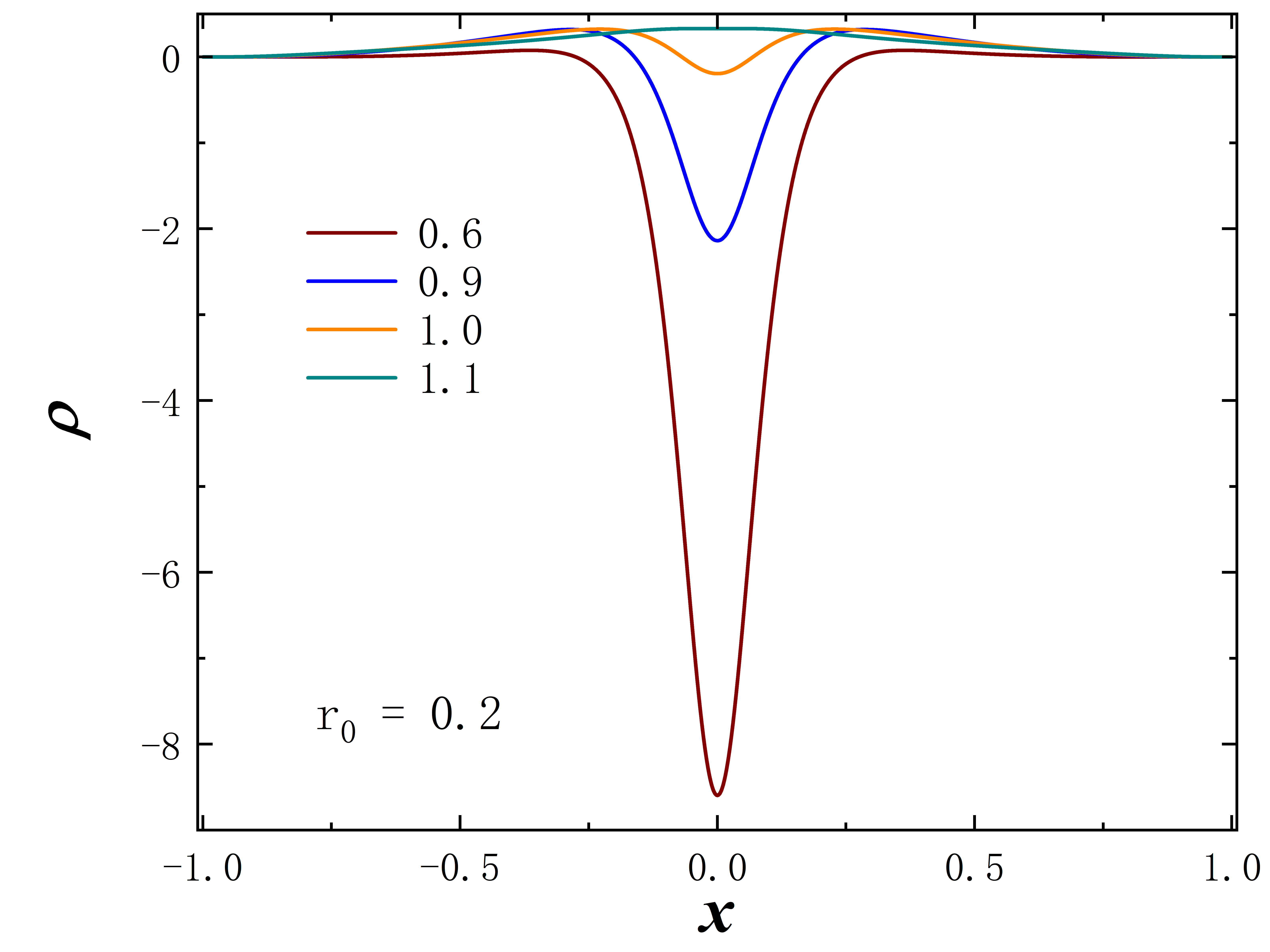}
\caption{Diverse combinations of the energy-tensor components are presented. In the first row, $r_0$ is set at 0.12 and in the second row, we choose $r_0 = 0.2$. The left figure selects varying $q$ values to illustrate the spatial distribution of $\rho + p_{i}$ and the solid line denotes $\rho + p_{1}$, the dotted line signifies $\rho + p_{2}$. The picture on the right shows the distribution of $\rho$ in space.}
\label{phase3}
\end{figure}

The traversable wormholes, which possess a phantom field with negative energy density, typically violate the Null Energy Condition (NEC) and other energy conditions. Upon introducing the magnetic monopole, which is contributed by the nonlinear electromagnetic field, we discovered some new phenomena. Similarly, we select $r_0$ as 0.12 and 0.2 and choose various values of $q$ as depicted in Fig. \ref{phase3}, considering that both positive and negative mass solutions exist for these parameters. For $r_0 = 0.12$, we choose the first solution for display, the solid line denotes $\rho + p_{1}$, and the dotted line signifies $\rho + p_{2}$. Firstly, main energy conditions, including the Null Energy Condition (NEC), are violated. However, upon examining specific components of the energy-momentum tensor, we find that this solution is only a minimal violation of NEC. For all solutions with $r_0 = 0.12, 0.2$,  the value of $\rho + p_{1}$ is less than 0, whereas for $\rho + p_{2}$, all solutions give results greater than 0 (this result holds for all solutions with different parameters $r_0$). Of particular interest is the alteration in the solution's total energy density. When $r_0 = 0.2$, the solution corresponds to ADM masses that are all positive, we observe that $\rho$ is less than 0 for small $q$ values, as the magnetic monopole's contribution is outweighed by the phantom field at this juncture. As $q$ increases, $\rho$ incrementally rises, and upon $q$ surpassing a critical threshold, $\rho$ exceeds 0 throughout the entire space. Moreover, when addressing the negative mass domain with $r_0 = 0.12$, an inverse pattern appears. The distribution of $\rho$ is positive at the throat of the wormhole, but farther on both sides of the throat, $\rho$ is negative(the exact position is different in different $q$). As $q$ increases, the values of $\rho$ near the throat do not change significantly. However, a substantial negative energy density emerges at greater distances in the universe on both sides of the wormhole.

To investigate the geometric characteristics of the wormhole, we select the parameter $r_0$ to be 0.12. Here $q = 1.33$ corresponds to the positive and negative mass branches of the first solution, and $q = 1.55$ corresponds to the negative mass in the second solution. By constructing the embeddings of planes with  $\theta = \pi/2$, and then using the cylindrical coordinates $(\rho,\varphi,z)$, the metric on this plane can be expressed by the following formula
\begin{align}
ds^2 &= C e^{-B}  d r^2 + C e^{-B} h   d\varphi^2 \, \\
&= d \rho^2 + dz^2 + \rho^2 d \varphi^2   \,.
\end{align}
Comparing the two equations above, we then obtain the expression for $\rho$ and $z$,
\begin{equation} \label{formula_embedding}
 \rho(r)= \sqrt{ C(r) e^{-B(r)} h(r) } ,
\end{equation}
\begin{equation}
z(r) = \pm  \int  \sqrt{ C(r) e^{-B(r)}  -   \left( \frac{d \rho}{d r} \right)^2    }     d r \;.
\end{equation}
Here $\rho$ corresponds to the circumferential radius, which corresponds to the radius of a circle located in the equatorial plane and having a constant coordinate $r$. The function $\rho(r)$ has extreme points, where the first derivative is zero. When the second derivative of the
extreme point is greater than zero, we call this point a throat. When the second derivative of the extreme point is less than zero, we call
this point an equator, which corresponds to a maximal surface.

\begin{figure}[htbp]
\centering
\includegraphics[width=0.9\linewidth]{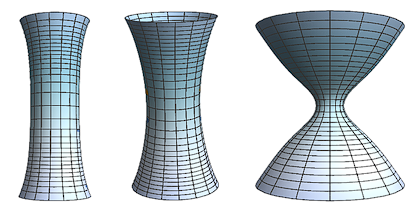}
\caption{The embedded image with throat size $r_0 = 0.12$ and $q = 2.5$ ($ M = 4.6614$), $q = 1.325$ ($ M = - 4.6614$) in the first solution, and $q = 1.55$ ($ M = - 3.5343$) in the second solution.}
\label{phase4}
\end{figure}

We show the embedding diagram in Fig. \ref{phase4}, from left to right, the image corresponds to $q = 2.5$ ($ M = 4.6614$), $q = 1.325$ ($ M = - 4.6614$) in the first solution, and $q = 1.55$ ($ M = - 3.5343$) in the second solution. Wormholes are always symmetrical and have only one throat and no equatorial plane.
\section{Conclusions}
The negative ADM mass that we currently derive depends on a specific model that introduces both an ``exotic'' phantom field and an untested magnetic monopole. Due to the need for the grand unified theory, it was hypothesized that magnetic monopoles should naturally occur in the universe. In the \cite{Gross:1983hb, LaCamera:2003zd}, investigators examined the gravitational implications of magnetic monopoles within specific models, and one finds that specific global magnetic monopole models provide negative effective mass \cite{Harari:1990cz}. On the other hand, there is no need for excessive concern regarding the ``exotic'' that arises from violating energy conditions. Back in the 1960s, the Trace Energy Condition (TEC), which stipulates that the trace of the energy-momentum tensor is non-negative, was entirely abandoned when it was realized that the equation of state it required does not apply to neutron stars. In accordance with the acceptance of observational data on the universe's accelerated expansion, the Strong Energy Condition (SEC) has been violated on a cosmological scale. The pressing need for the unification of quantum theory and gravity theory further suggests potential jeopardy for the Weak Energy Condition (WEC) and Null Energy Condition (NEC) \cite{Barcelo:2002bv}.

Before this, regardless of whether the traversable wormholes only with phantom fields or if the model was expanded to include additional matter fields \cite{Blazquez-Salcedo:2020czn, Konoplya:2021hsm, Chew:2016epf}, or even extended to some modification of gravity theory \cite{Bronnikov:2002rn, Harko:2013yb}, the creation of negative energy densities ensued, resulting in the violation of a series of energy conditions but no negative ADM mass was obtained. In our model, the occurrence of negative mass material coincides with a blue shift near the wormhole's throat, which may be attributed to the accumulation of a large negative energy density from the phantom field. From another perspective, the wormholes in the presence of negative mass can be understood as special white holes. The ejection of matter provides the negative sign in front of the numerical value and causes a blueshift to occur. This suggests the potential for a large-scale spatial distribution of negative mass matter, which could act as dark energy and exert repulsive effects in the universe. Perhaps, as Einstein said: ``\emph{a modification of the theory is required such that empty space takes the role of gravitating negative masses which are distributed all over the interstellar space}''.

This work is supported by the National Key Research and
Development Program of China (Grant No. 2020YFC2201503) and the National Natural
Science Foundation of China (Grant No.12047501 and No.12275110).

\end{document}